\documentclass[prb, showpacs, twocolumn, floatfix]{revtex4}
\usepackage{graphicx}
\usepackage{epstopdf}
\usepackage{amsmath, amsfonts, amssymb, bm}
\usepackage{subfigure}
\begin{document}
%%%%%%%%%%%%%%%%%%%%%%%%%%%%%%%%%%%%%%%%%%%%%%%%%%%%%%%%%%%%%%%%%%%%%%%%%%%%%%%%%%%%%%%%%%%%%%%%%%%%%%%%%%%%%%%%
\title{Collective quantum dot inversion and amplification of photon and phonon waves}
%%%%%%%%%%%%%%%%%%%%%%%%%%%%%%%%%%%%%%%%%%%%%%%%%%%%%%%%%%%%%%%%%%%%%%%%%%%%%%%%%%%%%%%%%%%%%%%%%%%%%%%%%%%%%%%%
\author{Sumanta \surname{Das}}
\email{Sumanta.Das@mpi-hd.mpg.de}
\author{Mihai A. \surname{Macovei}}
\email{macovei@phys.asm.md}
\affiliation{Institute of Applied Physics, Academiei str. 5, MD-2028 Chi\text{\c s}in\text{\u a}u, Moldova\\
Max-Planck-Institut f\"ur Kernphysik, Saupfercheckweg 1, D-69117 Heidelberg, Germany}
\date{\today}
%%%%%%%%%%%%%%%%%%%%%%%%%%%%%%%%%%%%%%%%%%%%%%%%%%%%%%%%%%%%%%%%%%%%%%%%%%%%%%%%%%%%%%%%%%%%%%%%%%%%%%%%%%%%%%%%
\begin{abstract}
The possibility of steady-state population inversion in a small sample of strongly driven 
two-level emitters like quantum dots (QDs) in micro-cavities, and its utilization towards amplification of light and 
acoustic waves is investigated theoretically. We find that inversion and absorption spectrum of photons, and phonons 
crucially depend on the interplay between the intrinsic vacuum and phonon environments. The absorption 
profiles of photons and phonons show marked novel features like gain instead of transparency and absorption 
reversed to gain, respectively. Furthermore, we report collectivity induced substantial enhancement of   
inversion and pronounced gain in the photon, and phonon absorption spectrum for a sub-wavelength-sized QD ensemble. 
\end{abstract}
%%%%%%%%%%%%%%%%%%%%%%%%%%%%%%%%%%%%%%%%%%%%%%%%%%%%%%%%%%%%%%%%%%%%%%%%%%%%%%%%%%%%%%%%%%%%%%%%%%%%%%%%%%%%%%%%
%%%%%%%%%%%%%%%%%%%%%%%%%%%%%%%%%%%%%%%%%%%%%%%%%%%%%%%%%%%%%%%%%%%%%%%%%%%%%%%%%%%%%%%%%%%%%%%%%%%%%%%%%%%%%%%%
\maketitle
%%%%%%%%%%%%%%%%%%%%%%%%%%%%%%%%%%%%%%%%%%%%%%%%%%%%%%%%%%%%%%%%%%%%%%%%%%%%%%%%%%%%%%%%%%%%%%%%%%%%%%%%%%%%%%%%

%%%%%%%%%%%%%%%%%%%%%%%%%%%%%%%%%%%%%%%%%%%%%%%%%%%%%%%%%%%%%%%%%%%%%%%%%%%%%%%%%%%%%%%%%%%%%%%%%%%%%%%%%%%%%%%%
\section{Introduction}
%%%%%%%%%%%%%%%%%%%%%%%%%%%%%%%%%%%%%%%%%%%%%%%%%%%%%%%%%%%%%%%%%%%%%%%%%%%%%%%%%%%%%%%%%%%%%%%%%%%%%%%%%%%%%%%%
Even after fifty years of the advent of laser, achieving population inversion and 
lasing in novel systems remain a topic of continuing interest \cite{Milb}. In particular, 
inversion in solid-state emitters like quantum dots (QDs) has
attracted considerable interests recently \cite{inv, exp_inv, inv_arp, Hug_PRL11}, due to the possibility of 
lasing in a driven two-level system \cite{sav,jon_PRL97} contrary to the common notion 
of multilevel manifolds \cite{Milb}. Semiconductor QDs are novel nano-structures 
with unique engineerable features like large optical dipole moments and  
emission wavelengths.  They can also be integrated efficient to micro-cavities and 
waveguides \cite{qdprop}. Their similarities to two-level atoms have been demonstrated 
in phenomena like Autler-Townes doublet \cite{Xu_sci}, Mollow triplet 
in resonance fluorescence \cite{Flag_nat,Vam_nat}, spectral line narrowing \cite{kei_PRL99,Oct_PRL,Matt_PRL12} 
and superradiance \cite{sup}. Furthermore, phonon-assisted excitation transfer in quantum dot molecules has been investigated in 
\cite{coll}. Additonally, due to potential application as 
solid state qubits numerous coherent optical studies of this system has been undertaken in 
the past decade \cite{qis,eit, ent1, pas, Awsch_nat08,Das_PRA11}.  
During last few years, sensitivity of QDs to the nature of their phonon environment has 
been exploited in numerous investigations. Phonon-assisted damping of Rabi oscillations \cite{size, dro,mon,Naz}, 
effect of phonons on polarization-entangled photons \cite{ent3}, phonon induced
transitions of excitons to cavity photons \cite{ph1}, heat pumping with optically 
driven excitons in phonon environment \cite{ph2} and effect of electron-phonon 
coupling on resonance fluorescence spectrum \cite{Ulrich_PRL11, res_f, Ulhaq} 
have been reported. Moreover, phonon-mediated population inversion in a single 
QD-cavity system driven by a continuous wave \cite{Hug2} and a pulse laser 
excitation \cite{Rei} was proposed recently. 

\begin{figure}[b]
\centering
\includegraphics[width=7.5cm]{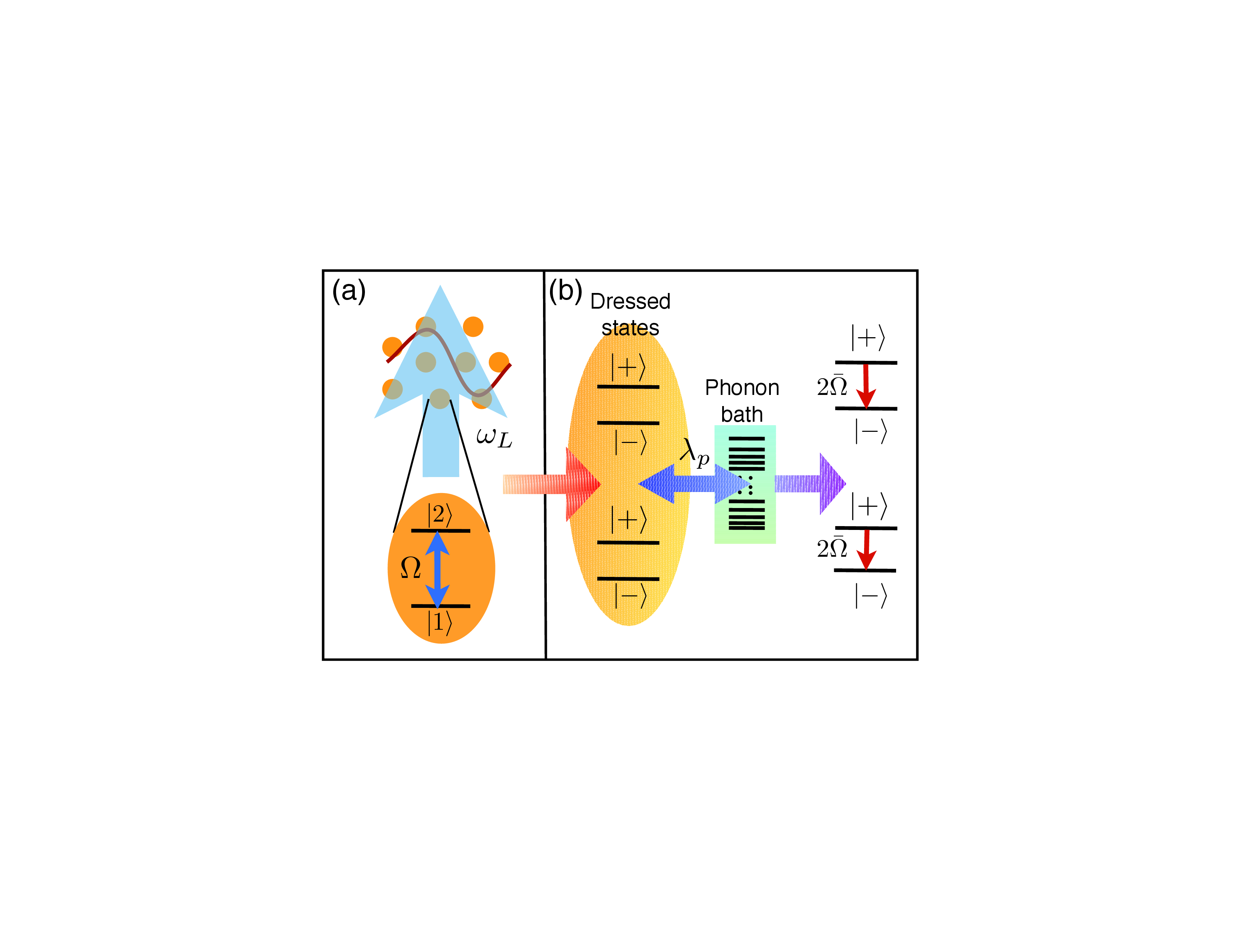}
\caption{\label{fig1}(color online) Schematics of interaction of a collection of 
QDs with a coherent field and phonon reservoir. (a) QDs
as two-level systems embedded in a substrate or in a micro-cavity of emission wavelength 
size, driven by a laser of Rabi frequency $\Omega$. (b) In strong field, 
the QDs are dressed by the laser giving rise to dressed-states $|\pm\rangle$
and characteristic Mollow triplet in emitted photon radiation. However, in presence of phonons,
additional transitions at the Rabi frequency $2\bar{\Omega}$ among the dressed-states 
are induced.}
\end{figure}

In light of these works, here, we propose a novel scheme to achieve and enhance 
population inversion in an ensemble of strongly driven two-level solid-state emitters 
like QDs via phonon-photon mediated collectivity (see Fig.~\ref{fig1}). We find 
important interplay of the phonon-photon reservoirs on the QD-dynamics and in creation 
of steady-state inversion in both bare- and dressed-states. As a key finding, we report 
phonon-induced unique features in the QD photon absorption spectrum like, gain instead 
of transparency at resonance and absorption switched to amplification for bare-state 
inversion. We also find amplification in the phonon absorption spectrum under condition 
of dressed-state inversion.
%%%%%%%%%%%%%%%%%%%%%%%%%%%%%%%%%%%%%%%%%%%%%%%%%%%%%%
%%%%%%%%%%%%%%%%%%%%%%%%%%%%%%%%%%%%%%%%%%%%%%%%%%%%%%%%%%%%%%%%%%%%
Furthermore, for a single QD, the phonon emission exhibits anti-bunched phonon 
statistics, i.e. a quantum effect. Additionally, for a sub-wavelength-sized 
ensemble of QDs, we report phonon-assisted collective enhancement of inversion 
and improvement of gain in the photon or phonon absorption spectra.

The article is organized as follows. In Section II, we describe the analytical 
approach and the system of interest, and obtain the master equation characterising 
the photon-phonon mediated interaction among a collection of two-level pumped 
qubits. In Section III, we analyze the obtained results, respectively. The Summary 
is given in Section IV.

%%%%%%%%%%%%%%%%%%%%%%%%%%%%%%%%%%%%%%%%%%%%%%%%%%%%%%%%%%%%%%%%%%%%%%%%%%%%%%%%%%%%%%%%%%%%%%%%%%%%%%%%%%%%%%%%
\section{System of interest}
%%%%%%%%%%%%%%%%%%%%%%%%%%%%%%%%%%%%%%%%%%%%%%%%%%%%%%%%%%%%%%%%%%%%%%%%%%%%%%%%%%%%%%%%%%%%%%%%%%%%%%%%%%%%%%%%
We consider an ensemble of QDs embedded in 
a substrate (such as InGaAs encased in GaAs substrate) or in a micro-cavity 
(like that formed by InAs/GaAs Bragg reflectors) illuminated by a cw laser of 
frequency $\omega_{L}$. As shown schematically in Fig.~(\ref{fig1}) the ground state 
$|1\rangle$ and the state containing a trapped electron-hole pair (exciton) $|2\rangle$ 
in each QD form a two-level system with transition frequency $\omega_{x}$. 
The linear dimension of the ensemble is smaller or of the order of the relevant emission wavelength. 
The laser is detuned from the exciton transition by $\Delta = \omega_{x}-\omega_{L}$ and drives it 
with a constant Rabi frequency of $\Omega$. Furthermore, the QDs interact with the solid state environment 
(phonon reservoir) and with field modes of the electromagnetic vacuum (photon reservoir) 
leading to different incoherent decay processes. The Hamiltonian describing the interaction of $N$ QDs with the laser and the vacuum and phonon reservoirs 
in a frame rotating with the laser frequency $\omega_{L}$  can be represented as:
%%%%%%%%%%%%%%%%%%%%%%%%%%%%%%%%%%%%%%%%%%%%%%%%%%%%%%%%%%%%%%%%%%
\begin{eqnarray}
\label{eq1}
\mathcal{H} & = &\sum_{k} \hbar \delta_{k}a^{\dagger}_{k}a_{k}+\sum_{p} \hbar \omega_{p}b^{\dagger}_{p}b_{p}+ \hbar \sum^{N}_{j = 1}\Delta S^{z}_{j}\nonumber\\
& + &\hbar \sum^{N}_{j = 1}\Omega~(S^{+}_{j}+S^{-}_{j}) + \big\{i\sum^{N}_{j = 1}\big[\sum_{p}\lambda_{p}S^{+}_{j}S^{-}_{j}b^{\dagger}_{p}\nonumber\\
& + &\sum_{k}(\vec{g}_{k}\cdot\vec{\wp}_{j})a^{\dagger}_{k}S^{-}_{j}\big]+ H.c.\big\}.
\end{eqnarray}
%%%%%%%%%%%%%%%%%%%%%%%%%%%%%%%%%%%%%%%%%%%%%%%%%%%%%%%%%%%%%%%%%%%
In Eq.~(\ref{eq1}) the first three terms represent the free Hamiltonians of the electromagnetic vacuum, the phonon reservoir and the two-level QDs.
The fourth term corresponds to the laser-QD interaction, while the two terms inside the curly brackets describe, respectively, the QDs interaction with the
surrounding phonon and vacuum reservoirs. The photon/phonon operators satisfy the standard 
boson commutation relations: $[Q_{u},Q^{\dagger}_{u'}]=\delta_{uu'}$ and $[Q_{u},Q_{u'}] = [Q^{\dagger}_{u},Q^{\dagger}_{u'}] = 0,
~ (Q = \{a, b\}, u =  \{k, p\})$. The terms $\lambda_{p}$, $\vec{g}_{k}$ and $\vec{\wp}_{j}$ are, respectively, the QD-phonon coupling constant taken 
in a general form satisfying the weak-coupling limit, the mode function of the three-dimensional multi-mode field, and the dipole moment of the $j$-th QD, 
while $\delta_{k} = \omega_{k}-\omega_{L}$. The raising and lowering operators for each QD are denoted by: $S^{+}_{j}=|2\rangle_{j}{}_{j}\langle 1|$,
$S^{-}_{j} = (S^{+}_{j})^{\dagger}$ and $S^{z}_{j} = (|2\rangle_{j}{}_{j}\langle 2|-|1\rangle_{j}{}_{j}\langle 1|)/2$ while
obeying the standard su($2$) angular momentum commutation relations.
In the following, we consider that the QDs system is driven by an intense laser.
This leads to the dressing of both the bare-states in each QD creating the  
dressed-states:
%%%%%%%%%%%%%%%%%%%%%%%%%%%%%%%%%%%%%%%%%%%%%%%%%%%%%%%%%%%%%%%%%%
\begin{eqnarray}
|+\rangle_{j} &=& \sin\theta |1\rangle_{j} +\cos\theta |2\rangle_{j}, \nonumber \\
|-\rangle_{j} &=& \cos\theta |1\rangle_{j} -\sin\theta |2\rangle_{j}. \label{dst}
\end{eqnarray}
%%%%%%%%%%%%%%%%%%%%%%%%%%%%%%%%%%%%%%%%%%%%%%%%%%%%%%%%%%%%%%%%%%
where $\tan 2\theta = (2\Omega/\Delta)$. Hence, it is imperative to continue the 
further analysis in the dressed-state basis.

The general form of the dressed master equation (DME) in the interaction picture 
is given by \cite{agab}: 
%%%%%%%%%%%%%%%%%%%%%%%%%%%%%%%%%%%%%%%%%%%%%%%%%%%%%%%%%%%%%%%%%%%%%%%%%%%%
\begin{eqnarray}
\partial \rho_{sb}/\partial t &=& 1/i\hbar\left[ \mathcal{H}_{I}(t), \rho_{sb}(0)\right] \nonumber \\
&-& 1/\hbar^{2}\int^{t}_{0}dt'
\left[ \mathcal{H}_{I}(t),\left[\mathcal{H}_{I}(t'), \rho_{sb}(t')\right] \right].
\end{eqnarray}
%%%%%%%%%%%%%%%%%%%%%%%%%%%%%%%%%%%%%%%%%%%%%%%%%%%%%%%%%%%%%%%%%%%%%%%%%%%%
Here, $\rho_{sb}$ is the density operator of the combined QDs $+$ reservoirs 
(both vacuum  and phonons) in the interaction picture. 
The dressed-state interaction Hamiltonian $\mathcal{H}_{I}$ can be evaluated by 
using the dressed-state transformations (\ref{dst}) in Eq.~(\ref{eq1}) and   
an unitary transformation by the operator $U=e^{(i/\hbar )H_{0}t}$, where 
%%%%%%%%%%%%%%%%%%%%%%%%%%%%%%%%%%%%%%%%%%%%%%%%%%%%%%%%%%%%%%%%%%%%%%%%%%%%
\begin{eqnarray}
H_{0} = \sum_{k}\hbar\delta_{k}a^{\dagger}_{k}a_{k}+\sum_{p} \hbar 
\omega_{p}b^{\dagger}_{p}b_{p}+ \hbar \bar{\Omega} \sum^{N}_{j = 1}R^{z}_{j}. 
\nonumber
\end{eqnarray}
%%%%%%%%%%%%%%%%%%%%%%%%%%%%%%%%%%%%%%%%%%%%%%%%%%%%%%%%%%%%%%%%%%%%%%%%%%%%
On substituting the interaction Hamiltonian and taking trace over the vacuum 
field and phonon modes, after some tedious algebra, we get the DME for the 
reduced density operator $\rho$ of the QDs in the Born-Markov and secular 
approximations, and weak coupling regime as: 
%%%%%%%%%%%%%%%%%%%%%%%%%%%%%%%%%%%%%%%%%%%%%%%%%%%%%%%%%%%%%%%%%%%%%%%%%%%%
\begin{eqnarray}
\label{eq6}
\frac{\partial \rho}{\partial t} &+& i\tilde{\Omega}\sum_{j}
\left[R^{z}_{j},\rho\right] =-\sum_{l,j}
\bigl \{ \chi_{0}\left[R^{z}_{l},R^{z}_{j}\rho\right] \nonumber\\
&+& \chi_{+}\left[R^{+}_{l},R^{-}_{j}\rho\right] + 
\chi_{-}\left[R^{-}_{l},R^{+}_{j}\rho\right] \bigr \} +  H.c., 
\end{eqnarray}
%%%%%%%%%%%%%%%%%%%%%%%%%%%%%%%%%%%%%%%%%%%%%%%%%%%%%%%%%%%%%%%%%%%%%%%%%%%%
where $\tilde{\Omega} =\bar{\Omega}-\Delta_{p}$ with 
$\bar{\Omega} = \sqrt{(\Delta/2)^{2}+\Omega^{2}}$ being 
the generalized Rabi frequency, while $\Delta_{p}$ is an 
additional shift due to QD-phonon coupling. The dressed basis operators are defined as:
$R^{z}_{j} = |+\rangle_{j j}\langle+|-|-\rangle_{j j}\langle-|$ and
$R^{+}_{j}  = |+\rangle_{j j}\langle-|, R^{-}_{j} = (R^{+}_{j} )^{\dagger}$
obeying the commutation relations: $[R^{+}_{j},R^{-}_{l}]=R^{z}_{j}\delta_{jl}$ and 
$[R^{z}_{j},R^{\pm}_{l}]=\pm 2 R^{\pm}_{j}\delta_{jl}$.

In Eq.~(\ref{eq6}), $\chi_{0,\pm}$ are the renormalized decay rates given by: 
%%%%%%%%%%%%%%%%%%%%%%%%%%%%%%%%%%%%%%%%%%%%%%%%%%%%%%%%%%%%%%%%%%%%%%%%%%%%
\begin{eqnarray}
\chi_{0} &=& \left(\Gamma_{0}\sin^{2}2\theta+\Gamma_{d}\cos^{2}2\theta\right)/4, \nonumber \\
\chi_{+} &=& \left[(\bar{n}+1)\Gamma_{p}+\Gamma_{d}\right]\sin^{2}2\theta/4 
+ \Gamma_{+}\cos^{4}\theta, \nonumber \\
\chi_{-} &=& \left(\bar{n}\Gamma_{p} +\Gamma_{d}\right)\sin^{2}2\theta/4+\Gamma_{-}\sin^{4}\theta, \nonumber
\end{eqnarray}
%%%%%%%%%%%%%%%%%%%%%%%%%%%%%%%%%%%%%%%%%%%%%%%%%%%%%%%%%%%%%%%%%%%%%%%%%%%%
respectively.
Here, $\Gamma_{0} = (\mathcal{G}/\hbar)^{2}\kappa/[\kappa^{2}+\delta^{2}_{c}] $, 
$\Gamma_{\pm} = (\mathcal{G}/\hbar)^{2}\kappa/[\kappa^{2}+(\delta_{c}\mp2\bar{\Omega})^{2}]$, 
$\delta_{c} = \omega_{c}-\omega_{L}$
and $\Gamma_{p} \equiv \Gamma_{p}(2\bar \Omega) = 
\pi\sum_{p}\left(\lambda_{p}/\hbar\right)^{2}\delta(\omega_{p}-2\bar\Omega)$ 
with $\mathcal{G} = (\vec{g}\cdot \vec{\wp})$, while 
$\sin2\theta = \Omega/\bar{\Omega}$, $\cos^{2}\theta = (\bar{\Omega}+\Delta/2)/2\bar{\Omega}$, 
and $\sin^{2}\theta = (\bar{\Omega}-\Delta/2)/2\bar{\Omega}$. Further, 
$\bar n \equiv \bar n(2\bar \Omega) = [\exp\left(2\hbar\bar \Omega/k_{B}T\right)-1]^{-1}$ 
is the mean-phonon number at frequency $2\bar \Omega$ and temperature $T$, with $k_{B}$ being the 
Bolzmann constant.
The vacuum-induced dressed decay rates $\Gamma_{0,\pm}$ are derived under the assumption that the QDs 
interact with a broad-band cavity mode of frequency $\omega_{c}$ and decay rate $\kappa$. 
$\Gamma_{p}$ is the phonon-induced dissipation rate of the QDs at the generalized Rabi frequency 
$2\bar{\Omega}$ (see Fig.~1b) whereas $\Gamma_{d}$ describes the dephasing rate of QDs \cite{ent3}. 
Notice that both reservoirs, i.e. photon and phonon, contribute to the collectivity among the QDs when 
fixed on a solid-state substrate. For this to occur, the inter-particle separations should be of the 
order of the smallest wavelength in the system (or less). Typically, the phonon wavelength in these 
samples is of the order of few/several nano-meters. Therefore, the Dicke limit considered here applies 
for a small QD system of an arbitrary shape or for specific geometries with moderate QDs numbers \cite{qis}.
Indeed, when two or more qubits are close to each other on the emission wavelength scale then collective 
effects occur, i.e. the photon/phonon emission of one qubit influences its neighbor qubit and vice-versa. 
A small-sized QDs ensemble can be well within the Dicke-limit with respect to photon emission wavelength. 
For phonons the situation is different. The phonon emission wavelength in our system (the phonon linear 
dispersion $\omega_{p} = c_{s}p$ is a good approximation in the long wave vector limit; here, $c_{s}$ is the 
speed of sound while $p$ is the phonon wave vector) is of the order of few/several $nm$ leading to 
inter-qubits phonon interactions only if the qubits's  separation is of this order or less. This can be 
achieved, for instance, in grid-like patterns as it was the case for experimentally observation of 
photon- induced superradiance with quantum dots \cite{sup} or in a quantum-dot array interacting via a 
common phonon bath \cite{qis}. Here, however, the Dicke limit with respect to phonon wavelength may not 
be satisfied meaning that the collective decay rates due to phonon emission will depend on the inter-qubit 
separations, $r_{jl}=|\vec r_{j}-\vec r_{l}|$. However, for a few/several-qubit ensemble  one can still 
consider that the collective decay rate is equal for all qubits. This happens because the inter-qubit 
separation in the collective decay rates enters as an argument to a sin/cos-like function \cite{qis}. 
Adjusting the inter-qubit separation with respect to the phonon wavelength one can avoid the inter-particle 
dependence of the phonon-induced decay rates for small-sized ensembles. Also, the exciton Bohr radius should 
be smaller than the size of a single QD, i.e. QDs are independent quantum objects in terms of 
quantum-mechanical tunnel coupling. Finally, in deriving Eq.~(\ref{eq6}), we have assumed that all dipole 
moments are equal, while for an inhomogeneous sample the frequency spread $\Delta\omega$ of the transition 
frequencies should be less than the collective decay rates, i.e. $\Delta \omega < N \chi_{0,\pm}$ 
(see, also, Ref.~\cite{sup}).

In the following, we shall investigate the photon-phonon mediated emission-absorption properties of an 
ensemble of two-level laser-pumped quantum dots described by Eq.~(\ref{eq6}).

%%%%%%%%%%%%%%%%%%%%%%%%%%%%%%%%%%%%%%%%%%%%%%%%%%%%%%%%%%%%%%%%%%%%%%%%%%%%%%%%%%%%%%%
\section{Results and discussion}
%%%%%%%%%%%%%%%%%%%%%%%%%%%%%%%%%%%%%%%%%%%%%%%%%%%%%%%%%%%%%%%%%%%%%%%%%%%%%%%%%%%%%%%
To understand the physics of QD dynamics, we next investigate the steady-state fluorescence 
and absorption spectrum of a single QD in the dressed basis and then generalize the 
obtained results to a multi-qubit ensemble. 

%%%%%%%%%%%%%%%%%%%%%%%%%%%%%%%%%%%%%%%%%%%%%%%%%%%%%%%%%%%%%%%%%%%%%%%%%%%%%%%%%
\subsection{Emission-absorption characteristics}
%%%%%%%%%%%%%%%%%%%%%%%%%%%%%%%%%%%%%%%%%%%%%%%%%%%%%%%%%%%%%%%%%%%%%%%%%%%%%%%%
The total steady-state photon fluorescence spectrum \cite{agab}, in bad cavity limit and 
under secular approximation is evaluated as:
$S_{\mathit{t}}(\omega) = S_{\mathit{c}}(\omega)+S(\omega)$, where 
$S_{\mathit{c}}(\omega)$ and $S(\omega)$ corresponds 
to the coherent and incoherent parts and for a single-qubit system are given by: 
%%%%%%%%%%%%%%%%%%%%%%%%%%%%%%%%%%%%%%%%%%%%%%%%%%%%%%%%%%%%%%%%%%%%%%%%%%%%%%%%
\begin{eqnarray}
\label{11}
S_{\mathit{c}}(\omega) = \frac{\pi}{4}\langle R^{z}\rangle^{2}_{ss}\sin^{2}2\theta\delta(\omega-\omega_{L}),
\end{eqnarray}
%%%%%%%%%%%%%%%%%%%%%%%%%%%%%%%%%%%%%%%%%%%%%%%%%%%%%%%%%%%%%%%%%%%%%%%%%%%%%%%%
and, respectively, 
%%%%%%%%%%%%%%%%%%%%%%%%%%%%%%%%%%%%%%%%%%%%%%%%%%%%%%%%%%%%%%%%%%%%%%%%%%%%%%%%
\begin{eqnarray}
\label{eq11a}
&{}&S(\omega)=\sin^{4}\theta\left(\frac{\chi_{+}}{\chi_{+}+\chi_{-}}\right)\frac{\Gamma_{ML}}{\Gamma^{2}_{ML}
+(\omega-\omega_{L}+2\tilde{\Omega})^{2}}\nonumber\\
&+&\sin^{2}2\theta\frac{\chi_{+}\chi_{-}}{\left(\chi_{+}+\chi_{-}\right)^{2}}\frac{\Gamma_{MC}}
{\Gamma^{2}_{MC}+(\omega-\omega_{L})^{2}}\nonumber\\
&+&\cos^{4}\theta\left(\frac{\chi_{-}}{\chi_{+}+\chi_{-}}\right)\frac{\Gamma_{MR}}
{\Gamma^{2}_{MR}+(\omega-\omega_{L}-2\tilde{\Omega})^{2}}.
\end{eqnarray}
%%%%%%%%%%%%%%%%%%%%%%%%%%%%%%%%%%%%%%%%%%%%%%%%%%%%%%%%%%%%%%%%%%%%%%%%%%%%%%%%%
%%%%%%%%%%%%%%%%%%%%%%%%%%%%%%%%%%%%%%%%%%%%%%%%%%%%%%%%%%%%%%%%%%%%%%%%%%%%%%%%
\begin{figure}[t]
\centering
\mbox{\subfigure{\includegraphics[width=4.1cm]{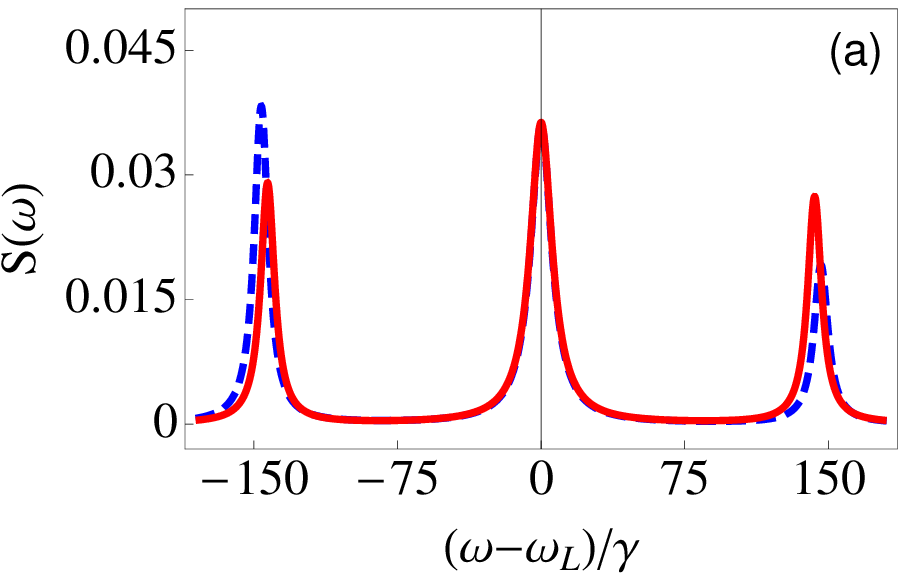}}
\subfigure{\includegraphics[width=4.4cm]{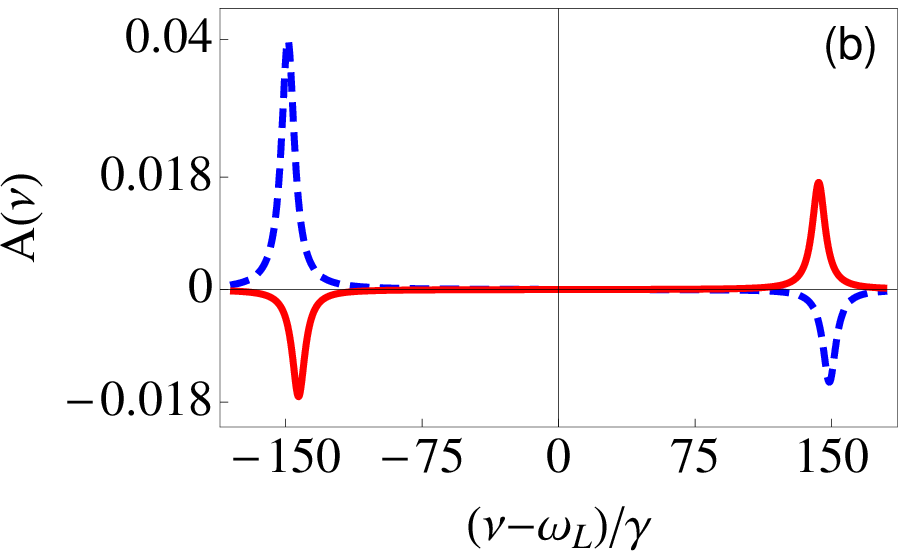} }}
\caption{ \label{fig2}(color online) The fluorescence and absorption spectrum of a 
QD at resonance (solid curve) and for negative detuning (dashed curve)
of $\Delta = -40$ $\mu$eV. The y-axis in both the plots are in dimensionless units and 
in (b) is multiplied by a factor of 10. 
We have considered parameters similar to that achievable in 
experiments: $\Omega = 80$ $\mu$eV, $\Delta_{p} = 20~\mu$eV, $\Gamma_{0} = \Gamma_{\pm} = \gamma = 0.84~\mu$eV, 
$\Gamma_{p} = 0.34~\mu$eV, $\Gamma_{d} = 3.84~\mu$eV and $T = 6$K. }
\end{figure}
%%%%%%%%%%%%%%%%%%%%%%%%%%%%%%%%%%%%%%%%%%%%%%%%%%%%%%%%%%%%%%%%%%%%%%%%%%%%%%%%
The corresponding photon absorption spectrum \cite{mac1} is: 
%%%%%%%%%%%%%%%%%%%%%%%%%%%%%%%%%%%%%%%%%%%%%%%%%%%%%%%%%%%%%%%%%%%%%%%%%%%%%%%%
\begin{eqnarray}
\label{eq11b}
A(\nu) & = & i\langle R^{z}\rangle_{ss}\bigg\{\sin^{4}\theta
\frac{\Gamma_{ML}+i(\nu-\omega_{L}+2\tilde{\Omega})}{\Gamma_{ML}^{2}+(\nu-\omega_{L}+2\tilde{\Omega})^{2}}\nonumber\\
& - &\cos^{4}\theta\frac{\Gamma_{MR}+i(\nu-\omega_{L}-2\tilde{\Omega})}
{\Gamma_{MR}^{2}+(\nu-\omega_{L}-2\tilde{\Omega})^{2}}\bigg\},
\end{eqnarray}
%%%%%%%%%%%%%%%%%%%%%%%%%%%%%%%%%%%%%%%%%%%%%%%%%%%%%%%%%%%%%%%%%%%%%%%%%%%%%%%%
while for phonons is given by:
%%%%%%%%%%%%%%%%%%%%%%%%%%%%%%%%%%%%%%%%%%%%%%%%%%%%%%%%%%%%%%%%%%%%%%%%%%%%%%%%
\begin{eqnarray}
\label{eq11c}
A(\nu_{p})  =  -\frac{i}{4}\langle R^{z}\rangle_{ss}\sin^{2}2\theta
\bigg\{\frac{\Gamma_{ML}+i(\nu_{p}-2\tilde{\Omega})}{\Gamma_{ML}^{2}+(\nu_{p}-2\tilde{\Omega})^{2}}\bigg\}.
\end{eqnarray}
%%%%%%%%%%%%%%%%%%%%%%%%%%%%%%%%%%%%%%%%%%%%%%%%%%%%%%%%%%%%%%%%%%%%%%%%%%%%%%%%
Here, $\langle R^{z}\rangle_{ss}$ is the single QD dressed basis steady-state 
population inversion given by: $(\chi_{-}-\chi_{+})/(\chi_{-}+\chi_{+})$.
$\Gamma_{ML} = \Gamma_{RL} = 4\chi_{0}+\chi_{+}+\chi_{-}$ and 
$\Gamma_{MC} = 2(\chi_{+}+\chi_{-})$ with $\{ \nu,\nu_{p}\}$ being the 
weak photon and phonon probe frequencies, respectively. 

In Fig.~\ref{fig2}(a) \& \ref{fig2}(b), we plot Eq.~(\ref{eq11a}) and imaginary part of Eq.~(\ref{eq11b}). 
The solid (dashed) lines in the plot corresponds to resonant (non-resonant) 
coherent excitation of the QD. 
The photon fluorescence spectrum shown in Fig.~\ref{fig2}(a) is similar to the Mollow spectrum well known for 
atoms, however, with certain key modifications due to the phonon bath.
The position of the Mollow sidebands are shifted to $\pm 2\tilde{\Omega}$ with phonon induced 
broadening of their widths $\Gamma_{ML}$ and that of the central line $\Gamma_{MC}$ as can be
easily seen from their respective analytical forms. 
Additionally, in presence of phonons the sidebands are asymmetric 
with the ratio of the right to left peak heights as: $[(2\bar{\Omega}+\Delta)/(2\bar{\Omega}-\Delta)]^{2}\chi_{-}/\chi_{+}$. 
This asymmetry is pronounced for non-resonant driving of the QD as shown in Fig.~\ref{fig2}(a) by the dashed curve. 
For resonant drive however, the asymmetry can be prominent in the 
conditions: low temperature reservoir and $\Gamma_{p} \gg (\Gamma_{d}, \gamma)$.
The coherent part of the fluorescence spectrum $S_{c}(\omega)$, that signifies the elastic scattering process in presence
of phonons, does not vanish even at resonance and has an amplitude $\propto \Gamma^{2}_{p}/[(2\bar{n}+1)\Gamma_{p}+2(\Gamma_{d}+\gamma)]^2$
(see, also, Ref.~\cite{coh_sc}). 
Note that, such phonon induced modifications of the Mollow triplet has been observed in a recent experiment \cite{Ulhaq}.

The imaginary part of the photon absorption spectrum shown in Fig.~\ref{fig2}(b) is strikingly different 
from a typical Mollow absorption spectrum of a two-level atom. As a key finding, for resonant excitation of a QD 
we report \textit{phonon induced amplification and absorption} in the spectrum at $\nu-\omega_{L} = \mp 2\tilde{\Omega}$ 
respectively with peak heights $\propto \langle R^{z}\rangle_{ss}/4$. 
Note that, these characteristics are absent in atomic system, where there is no phonon coupling and the probe field is 
transparent at resonance as $\langle R^{z}\rangle_{ss} = 0$. Furthermore,  
for both resonant and non-resonant excitation, we find from Eq.~(\ref{eq11b}) broadening of the amplification
and absorption spectral lines due to phonons. The imaginary part of Eq.~(\ref{eq11c}) exhibits another 
remarkable feature, \textit{gain} in the phonon absorption spectrum for \textit{population inversion in the dressed-state}. 
This, in principle, leads to amplification of a weak phonon wave in presence of an additional phonon cavity \cite{tri,Jul_PRL12} 
tuned to the dressed transition at the shifted generalized Rabi frequency of $2\tilde{\Omega}$ (see Fig.~1b). Note that, the 
phonon amplification profile is enhanced only in presence of the vacuum reservoir as can be seen from the expressions for 
$\chi_{0,\pm}$.

For a collection of $N$ QDs in a sub-wavelength-sized ensemble, the emission-absorption spectrum 
is modified due to collective effects. The spectral lines are further broadened and enhanced, with the enhancement proportional to $N^2$ 
and $N$, respectively, for the Mollow peaks in photon fluorescence and gain profile in the photon or phonon absorption \cite{mac1}. 
In the following, we discuss another key finding of the paper: \textit{collectivity assisted 
enhancement} of steady-state population inversion. 

%%%%%%%%%%%%%%%%%%%%%%%%%%%%%%%%%%%%%%%%%%%%%%%%%%%%%%%%%%%%%%%%%%%%%%%%%%%%%%%%%
\subsection{Enhanced inversion due to photon-phonon mediated collectivity}
%%%%%%%%%%%%%%%%%%%%%%%%%%%%%%%%%%%%%%%%%%%%%%%%%%%%%%%%%%%%%%%%%%%%%%%%%%%%%%%%
In order to study the collective steady-state population inversion of the QDs in 
bare- and dressed-basis, we need to obtain the steady-state solution of Eq.~(\ref{eq6}). 
For this purpose, we first define the dressed-state collective operators as: 
$\mathcal{R}^{\pm} = \sum^{N}_{j=1}R^{\pm}_{j}$, and $\mathcal{R}^{z} = \sum^{N}_{j=1}R^{z}_{j}$, 
and then look for a solution of the form:
%%%%%%%%%%%%%%%%%%%%%%%%%%%%%%%%%%%%%%%%%%%%%%%%%%%%%%%%%%%%%%%%%%%%%%%%%%%
\begin{eqnarray}
\label{eq12}
\rho_{ss} =Z^{-1}\exp[-\eta\mathcal{R}^{z}],
\end{eqnarray}
%%%%%%%%%%%%%%%%%%%%%%%%%%%%%%%%%%%%%%%%%%%%%%%%%%%%%%%%%%%%%%%%%%%%%%%%%%%
where the normalization $Z$ is determined by the requirement $\mathsf{Tr}\{\rho_{s}\} = 1$. 
The unknown variable $\eta$ can be evaluated by inserting Eq.~(\ref{eq12}) in the DME 
and taking the steady-state $d\rho/dt = 0$ condition. On doing so, we obtain:  
$\eta =\ln\left(\chi_{+}/\chi_{-}\right)/2$.
Note that, the coherent part of the dipole-dipole interaction potential commutes with the 
steady-state solution (\ref{eq12}) obtained for a small-sized ensemble meaning that 
it does not modify the population distribution (unless it is larger than the generalized 
Rabi frequency). This can be seen by adding an extra term to the Hamiltonian (\ref{eq1}),
namely:
%%%%%%%%%%%%%%%%%%%%%%%%%%%%%%%%%%%%%%%%%%%%%%%%%%%%%%%%%%%%%%%%%%%%%%%%%%%
\begin{eqnarray}
H_{dd} = \Omega_{dd}\sum^{N}_{j\not=l=1}S^{+}_{j}S^{-}_{l}, \label{Ddd}
\end{eqnarray}
%%%%%%%%%%%%%%%%%%%%%%%%%%%%%%%%%%%%%%%%%%%%%%%%%%%%%%%%%%%%%%%%%%%%%%%%%%%
where the averaged over all possible orientation of dipoles coherent dipole-dipole 
interaction behaves as: $\Omega_{dd} \propto 1/(2r_{jl})$, and it was taken identical 
for all involved qubits - an approximation valid for smaller ensembles. Applying the 
dressed-state transformation (\ref{dst}) to (\ref{Ddd}) in the secular approximation, 
one can observe that the Hamiltonian describing the dipole-dipole interaction includes 
the collective correlators: $R^{{+}}R^{{-}}$, $R^{{-}}R^{{+}}$ and $R^{2}_{z}$, respectively, 
commuting with the solution (\ref{eq12}). We next look for the inversion in the bare-states. 
In the secular approximation the collective bare-state inversion is given by:
%%%%%%%%%%%%%%%%%%%%%%%%%%%%%%%%%%%%%%%%%%%%%%%%%%%%%%%%%%%%%%%%%%%%%%%%%%%
\begin{eqnarray}
\label{eq14}
& &\langle\mathcal{S}^{z}\rangle_{ss} =\left(\Delta/2\bar \Omega\right)
\langle\mathcal{R}^{z}\rangle_{ss}/2,
\end{eqnarray}
%%%%%%%%%%%%%%%%%%%%%%%%%%%%%%%%%%%%%%%%%%%%%%%%%%%%%%%%%%%%%%%%%%%%%%%%%%%
where $\langle\mathcal{R}^{z}\rangle_{ss}$ is the collective steady-state 
dressed-state inversion found from Eq.~(\ref{eq12}) as \cite{jon_PRL97}:
\begin{equation}
\label{eq15}
\langle \mathcal{R}^{z}\rangle_{ss} =N\left(\frac{1+e^{2\eta(N+1)}}
{1-e^{2\eta(N+1)}}\right)+\frac{2(e^{2\eta(N+1)}-e^{2\eta})}{(1-e^{2\eta})(1-e^{2\eta(N+1)})}.
\end{equation}
To gain physical insight into the behavior of the bare-state population, we need to understand the 
inversion of dressed-state populations. For this purpose, we consider two limits in Eq.~(\ref{eq15}): 
(i) $e^{2\eta} = \xi = \chi_{+}/  \chi_{-} \ll 1$ and (ii) $\xi = \chi_{+}/  \chi_{-} \gg 1$.
From Eq.~(\ref{eq15}), we get in the limit (i): $\langle \mathcal{R}^{z}\rangle_{ss}/N \simeq 1- [2\xi/N(1-\xi)]$, 
which clearly approaches maximum inversion $\langle \mathcal{R}^{z}\rangle_{ss}/N  = 1$, for 
$N \gg 1$. As $N$ represent the number of QDs in the ensemble, we can conclude that collectivity 
among the QDs enhances population inversion in the dressed-state. 
In the limit (ii) on the other hand, we get: $\langle \mathcal{R}^{z}\rangle_{ss}/N \simeq [2/N(\xi-1)]-1$
which for $N \gg 1$ becomes, $\langle \mathcal{R}^{z}\rangle_{ss}/N = -1$, 
suggesting that all the population tends to reside in the lower dressed-state $|-\rangle$ and, thus, 
no inversion. Note that, in the limit $\chi_{+}/  \chi_{-} \rightarrow 1$, $\langle \mathcal{R}^{z}\rangle_{ss}/N \rightarrow 0$
there is also no inversion as the dressed-states are equally populated. 
Additionally, we would like to emphasize that inversion is possible 
\textit{only in presence of the radiative decay} $\Gamma_{\pm}$ 
\textit{among the dressed-states} with $\Gamma_{\pm} > \Gamma_{p}$. 
This can be proved by analyzing the expressions for $\chi_{\pm}$. 
On substituting $\Gamma_{\pm} = 0$ we have 
$\chi_{+}/ \chi_{-} =[(\bar{n}+1)\Gamma_{p}+\Gamma_{d}]/[\bar{n}\Gamma_{p}+\Gamma_{d}] \ge 1$, 
for any bath temperature and QD parameters thereby implying \textit{no possible dressed-state inversion}. 
Finally, setting $\{ \Gamma_{\pm}, \Gamma_{d}\} = 0$ and $N=1$, one can recover the thermal population 
distribution among the dressed-states \cite{th_d}. Indeed, one can easily show that the steady-state upper 
and lower dressed-state populations (see, also, Fig.~\ref{fig1}) are, respectively, 
$\langle R^{++}\rangle_{ss} = \bar n/(1+2\bar n)$ and $\langle R^{--}\rangle_{ss} = (1+\bar n)/(1+2\bar n)$. 
Their respective ratio is: $\langle R^{++}\rangle_{ss}/\langle R^{--}\rangle_{ss} = 
\bar n/(1+\bar n) \equiv e^{-\beta}$, where $\beta = \frac{2\hbar \bar \Omega}{k_{B}T}$, and, thus, behaves 
according to thermal equilibrium relaxation. This will be the case also if spontaneous emission due to 
interaction with the vacuum modes of the surrounding electromagnetic field reservoir is smaller than the 
phonon-induced decay rates.
%%%%%%%%%%%%%%%%%%%%%%%%%%%%%%%%%%%%%%%%%%%%%%%%%%%%%%%%%%%%%%%% 
\begin{figure}[!t]
\centering
\mbox{\subfigure{\includegraphics[width=4.28cm]{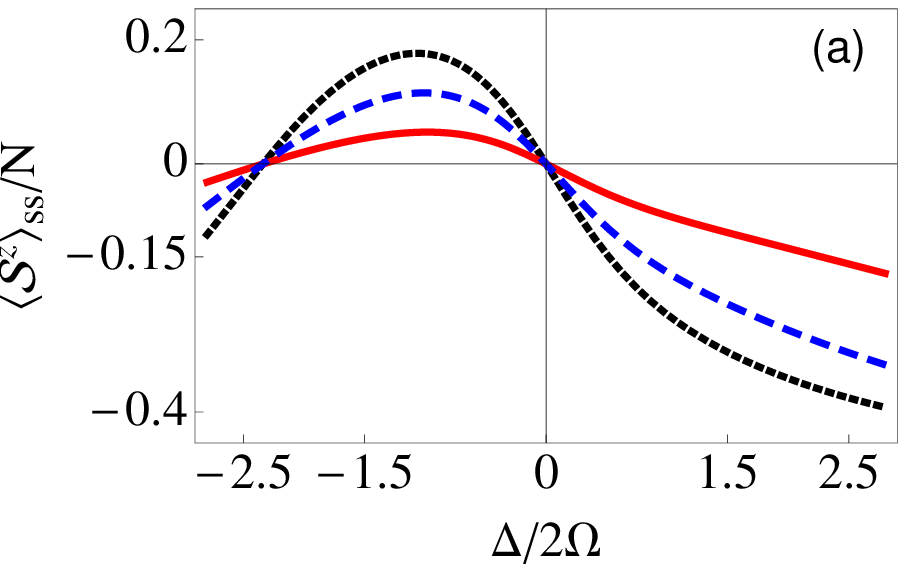}}
%\hspace{0.01cm}
\subfigure{\includegraphics[width=4.28cm]{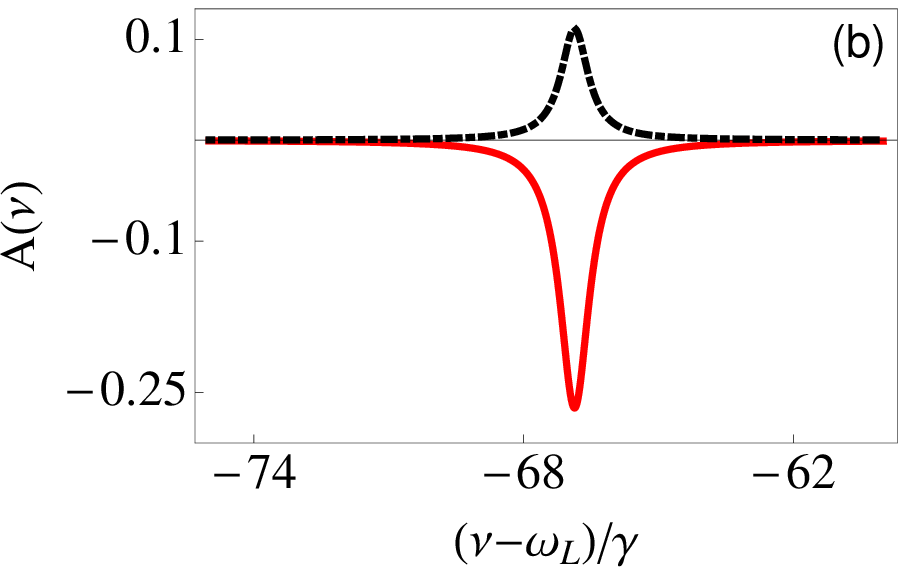} }}
\caption{ \label{fig3}(color online) (a) Steady-state population inversion in the bare-state
for $N=10$ (dotted curve), $N=5$ (dashed line) and $N=1$ (solid curve) QDs.  
(b) Absorption spectrum of a single QD for $\Delta/2\Omega = -1$ (maximum inversion in Fig.~3a) 
in presence (solid curve) and absence (dot-dashed line) of phonons. 
In (b) the y-axis is in dimensionless units and multiplied by a factor of $100$ 
while the x-axis by $0.01$.  Parameters for the plots are:
$\Omega = 2$~meV, $\Delta_{p} = 20~\mu$eV, $\Gamma_{0} = \Gamma_{\pm} = \gamma = 0.84~\mu$eV, 
$\Gamma_{p} = 20\gamma$, $\Gamma_{d} = 40\gamma$ and $T = 6$K.}
\end{figure}
%%%%%%%%%%%%%%%%%%%%%%%%%%%%%%%%%%%%%%%%%%%%%%%%%%%%%%%%%%%%%%%%
 
It is clear from the above discussion and Eqs~(\ref{eq14}) and (\ref{eq15}), that 
for conditions of inversion in the dressed-state there will be no 
population inversion among the bare-states $|1\rangle$ and $|2\rangle$. However, 
at low temperature (small $\bar{n}$), if $\Gamma_{p} \gg \Gamma_{\pm}$ we can satisfy the condition 
$\chi_{+}/\chi_{-} \gg 1$ and $\langle \mathcal{R}^{z}\rangle/N < 0$
thereby leading to inversion among the bare-states for negative laser detuning. 
In Fig.~\ref{fig3}(a), we show the steady-state population inversion among the bare-states for 
different number of QDs and typical experimental parameters \cite{ph2}. 
Inversion is seen to be achievable for certain range of 
negative laser detuning as expected from the above discussion. 
Additionally, we find collectivity induced enhancement of population inversion in the bare-state.
This follows directly from the behavior of $\langle \mathcal{R}^{z}\rangle_{ss}/N$ for 
$\chi_{+}/\chi_{-} \gg 1$. We report almost $40\%$ inversion for reasonable ($N = 10$) number of 
QDs (complete inversion in bare-state corresponds to a $\langle S^{z}\rangle/N = 0.5)$.
Note that the mean-phonon number depends on the generalized Rabi frequency and, thus, varies as detuning varies.
This has been explicitly considered here. In Fig.~\ref{fig3}(b), we show the behavior of the imaginary part 
of photon absorption spectrum of Eq.~(\ref{eq11b}) at maximum bare-state inversion for a single QD. The spectrum 
show completely new feature in presence of phonons - \textit{amplification at bare-state inversion}.
However, for $\Gamma_{\pm} \gg \Gamma_{p}$ or in absence of phonons, we get back the typical 
absorption peak instead of gain (dot-dashed line in Fig.~\ref{fig3}b).
This implies that we can get amplification of a probe laser from a QD ensemble. 

Thus, concluding, for a small-sized ensemble of QDs, we find that collectivity 
enhances the inversion substantially both in bare and dressed-state. Hence, the 
\textit{phonon induced} gain characteristics of photon-absorption spectrum 
that we report here for bare-state inversion, can be utilized towards amplification 
of a weak light field at optical frequencies in such solid-state systems. 
The \textit{photon induced} gain characteristic of the phonon-absorption 
spectrum for dressed-state inversion, on the other hand, 
can amplify a phonon wave at THz frequencies tuned to the $2\bar{\Omega}$ 
transition (as $\bar{\Omega}\sim$ ps is achievable in QDs). Furthermore, the 
phonon emission of a single pumped QD shows anti-bunched phonon statistics, namely: 
%%%%%%%%%%%%%%%%%%%%%%%%%%%%%%%%%%%%%%%%%%%%%%%%%%%%%%%%%%%%%%%%%%%
\begin{eqnarray}
g^{(2)}(\tau) = 1-\exp[-2(\chi_{+} + \chi_{-})\tau]. \label{g2}
\end{eqnarray}
%%%%%%%%%%%%%%%%%%%%%%%%%%%%%%%%%%%%%%%%%%%%%%%%%%%%%%%%%%%%%%%%%%%
This can be understood as follows: once a phonon is emitted - the qubit will be in the lower dressed-state. 
Obviously, the qubit will generate then a photon on the dressed-state transition $| - \rangle \to | + \rangle$ 
or $| - \rangle \to | - \rangle$, respectively (see Fig.~\ref{fig1}). Therefore, one can obtain in principle 
a flux of single-phonons. If the qubit is pumped with a strong pulse, then one can manage to obtain a 
single-phonon anti-bunched source. Here, one can look also at the photon-photon, phonon-phonon, phonon-photon 
or photon-phonon induced correlations. Thus, photon-phonon correlations open interesting regimes of 
quantum statistics in these systems which can be harnessed towards quantum computation.

%%%%%%%%%%%%%%%%%%%%%%%%%%%%%%%%%%%%%%%%%%%%%%%%%%%%%%%%%%%%%%%%%%%
\section{Summary}
%%%%%%%%%%%%%%%%%%%%%%%%%%%%%%%%%%%%%%%%%%%%%%%%%%%%%%%%%%%%%%%%%%%
Summarising,  we have derived a collective dressed-state master equation 
for an ensemble of strongly driven QDs coupled to phonon and vacuum reservoirs.  
As a key finding, we reported gain characteristic in the photon and phonon absorption 
spectrum instead of transparency and absorption respectively due to crucial interplay 
of the vacuum and phonon reservoirs. Furthermore, we have shown that phonon 
mediated collectivity among the QDs enhances the population inversion in the bare-states.  
Anti-bunching occurs for phonon emission by a single pumped QD. Our findings show the 
possibility for amplification of photon and phonon waves as well as quantum particle correlations, 
and can open new directions towards creation of nanoscale optical and acoustic 
quantum or classical coherent sources.  

%%%%%%%%%%%%%%%%%%%%%%%%%%%%%%%%%%%%%%%%%%%%%%%%%%%%%%%%%%%%%%%%%%%
\section{Acknowledgement}
%%%%%%%%%%%%%%%%%%%%%%%%%%%%%%%%%%%%%%%%%%%%%%%%%%%%%%%%%%%%%%%%%%%
We acknowledge insightful discussions with Prof.
C. H. Keitel. S.D. is grateful for the hospitality of the 
Institute of Applied Physics, Chi\text{\c s}in\text{\u a}u, Moldova, 
while M.M. acknowledges support via the research grant Nr. 13.820.05.07/GF.

%%%%%%%%%%%%%%%%%%%%%%%%%%%%%%%%%%%%%%%%%%%%%%%%%%%%%%%%%%%%%%%%%%%%

%%%%%%%%%%%%%%%%%%%%%%%%%%%%%%%%%%%%%%%%%%%%%%%%%%%%%%%%%%%%%%%%%%%%%%%%%%%%%%%%%%%%%%%%%%

\end{document}